\documentclass{article}

\usepackage{apj}

\received{}
\accepted{}

\begin{document}

\title{Radio Emission from 3D Relativistic Hydrodynamic Jets: Observational
Evidence of Jet Stratification}

\author{Miguel-Angel Aloy\altaffilmark{1}, 
Jos\'e-Luis G\'omez\altaffilmark{2}, 
Jos\'e-Mar\'{\i}a Ib\'a\~nez\altaffilmark{1},
Jos\'e-Mar\'{\i}a Mart\'{\i}\altaffilmark{1} and Ewald
M\"uller\altaffilmark{3}
}

\altaffiltext{1}{Departamento de Astronom\'{\i}a y Astrof\'{\i}sica,
Universidad de Valencia, 46100 Burjassot (Valencia), Spain.
miguel.a.aloy@uv.es; jose.m.ibanez@uv.es; Jose-Maria.Marti@uv.es}

\altaffiltext{2}{Instituto de Astrof\'{\i}sica de Andaluc\'{\i}a, CSIC,
Apartado 3004, 18080 Granada, Spain. jlgomez@iaa.es}

\altaffiltext{3}{Max-Planck-Institut f\"ur Astrophysik,
Karl-Schwarzschild-Str. 1, 85748 Garching, Germany. ewald@mpa-garching.mpg.de}

\begin{abstract}

  We present the first radio emission simulations from high resolution three
dimensional relativistic hydrodynamic jets, which allow for a study of the
observational implications of the interaction between the jet and external
medium. This interaction gives rise to a stratification of the jet where a
fast spine is surrounded by a slow high energy shear layer. The
stratification, and in particular the large specific internal energy and slow
flow in the shear layer largely determines the emission from the jet. If the
magnetic field in the shear layer becomes helical (e.g., resulting from an
initial toroidal field and an aligned field component generated by shear) the
emission shows a cross section asymmetry, in which either the top or the
bottom of the jet dominates the emission. This, as well as limb or spine
brightening, is a function of the viewing angle and flow velocity, and the
top/bottom jet emission predominance can be reversed if the jet changes
direction with respect to the observer, or presents a change in velocity. The
asymmetry is more prominent in the polarized flux, because of field
cancellation (or amplification) along the line of sight. Recent observations
of jet cross section emission asymmetries in the blazar 1055+018 can be
explained assuming the existence of a shear layer with a helical magnetic
field.


\end{abstract}

\section{Introduction}

  The development of high-resolution multidimensional relativistic
hydrodynamic codes has provided a tool which allows to simulate the
(synchrotron) radio emission from parsec-scale relativistic jets (G\'omez et
al. \cite{JL95}, \cite{JL97}; Komissarov \& Falle \cite{KF96}, \cite{KF97};
Mioduszewski, Hughes \& Duncan \cite{MI97}), obtaining a better understanding
of the physics involved in the jets of active galactic nuclei and their
enviroments. It has also been used to successfully explain the structure of
particular sources (e.g., 3C~120, G\'omez et al. 1998a,b; 3C~454.3, G\'omez,
Marscher \& Alberdi \cite{JL99}).

  In this Letter we study, for the first time, the radio emission properties
of three-dimensional relativistic hydrodynamic jet models. In particular, we
focus on the observational consequences of the interaction between the
relativistic jet and the surrounding medium, which leads to the development of
a shear layer. Such shear layers (with distinct kinematical properties and
magnetic field configuration) appear naturally in some models of jet formation
(Sol, Pelletier, \& Asseo \cite{SPA89}) and have been invoked in the past by
several authors (Komissarov \cite{K90}, Laing \cite{L96}, Laing et
al. \cite{L99}) in order to account for a number of observational
characteristics of FRI radio sources.  However, the physical nature of the
shear layer is still largely unknown. Recently, Swain, Bridle \& Baum
(\cite{SB98}) have found evidence of shear layers in FRII radio galaxies
(3C353), and Attridge, Roberts \& Wardle (\cite{ARW99}) have inferred a
two-component structure in the parsec scale jet of the source 1055+018.

\section{Jet Stratification: Beam and Shear Layer}

  To study the emission properties of relativistic jets we have used the high
resolution three dimensional relativistic hydrodynamic jet model of Aloy et
al. (\cite{MA99a}, hereafter A99). The model is characterized by a
beam-to-external proper rest-mass density ratio $\eta=0.01$, a beam Mach
number $M_b=6.0$, and a beam flow speed $v_b = 0.99c$ ($c$ is the speed of
light) corresponding to a beam Lorentz factor of $\Gamma \sim
7$. Non-axisymmetry was triggered by means of a helical velocity perturbation
of 1\% amplitude and with a period of 3.0 $R_b/c$ (where $R_b$ is the initial
beam radius) imposed at the nozzle. We refer the reader to Aloy et
al. (\cite{MA99b}) where a detailed description of the hydrodynamical code can
be found.

\begin{figure*}
\plotone{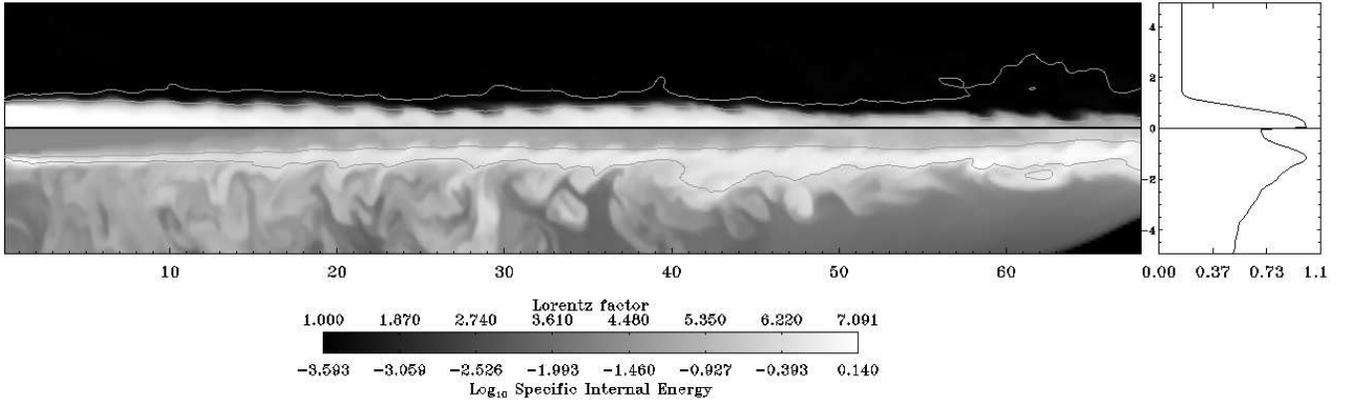}
%\vspace{4cm}
\caption{Cuts of the Lorentz factor (top half panel) and specific internal
energy (bottom half panel) distributions of the hydrodynamic model along the
plane $y=0$. White contours representing constant values of the beam particle
fraction (0.95 for the innermost contour, 0.2 for the outermost one) are used
to characterize the shear layer. The two panels at the right show the average
(along lines $x=$constant) of the corresponding distributions across the jet.}
\label{fig1}
\end{figure*}

  The jet model is characterized by a two-component structure
(Fig.\,\ref{fig1}) with a fast ($\Gamma \sim 7$) inner jet and a slower
($\Gamma \sim 1.7$) shear layer with high specific internal energy.  The shear
layer is defined as the region where the beam particle fraction is between 0.2
and 0.95. As discussed in A99, the formation of the shear layer is dominated
by the numerical viscosity inherent to the hydrodynamic code and not by the
turbulent shear. Inspite of this fact, the computed jet models still allow to
study the physics of shear layers in relativistic jets and their observational
consequences. As shown in A99, the axial component of the momentum of the beam
particles decreases by 30\% within the first 60 $R_b$. This loss of momentum
causes a decrease of the Lorentz factor in the inner jet with values $\sim$
5.8, 5.3, and 4.8 at $z=$ 25, 50, and 68 $R_b$, respectively.
  
\section{Emission properties}

  The emission properties of large scale jets in AGNs can be studied by
computing the radio (synchrotron) emission from relativistic hydrodynamic jet
models (see G\'omez et al. \cite{JL95},\cite{JL97} and references therein for
a complete description of the model). For this, we assume that the particle
and energy density of the non-thermal electrons is a constant fraction of the
simulated thermal gas. No significant variations from this proportionality are
expected to be found within the jet, since the radiative loses and particle
accelerations in our model are small. The internal energy among the
relativistic non-thermal electrons is distributed following a power law. The
magnetic energy is set to be locally proportional and significantly smaller
than the particle energy density, hence being dynamically
negligible. Different {\it ad hoc} distributions of the magnetic field in the
jet spine and shear layer can therefore be considered. In our model we assume
that the magnetic field of the jet consists of two components. A toroidal
field present both in the jet spine and the shear layer, and a second
component (in equipartition with the toroidal field) aligned in the shear
layer and radial in the jet spine. The aligned component in the shear layer
could arise from the shear between the jet and the external medium, while the
radial field in the jet spine may be due to transverse shocks (i.e., Attridge,
Roberts \& Wardle \cite{ARW99}). The resulting projected magnetic field is
aligned in the shear layer and is perpendicular in the jet spine, as suggested
by several observations (Laing \cite{L96}; Swain, Bridle \& Baum \cite{SB98};
Attridge, Roberts \& Wardle \cite{ARW99}). An extra randomly oriented magnetic
field component (containing 60\% of the total magnetic field energy) is
assumed both for the shear layer and jet spine.

  The Stokes parameters that determine the emission are calculated by
integrating the synchrotron transfer equations along columns parallel to the
line of sight accounting for the appropriate relativistic effects, such as
Doppler boosting and light aberration. Light--travel time delays have been
ignored assuming that the jet is stationary. Since only the jet material is
expected to radiate, the energy density computed with the hydro code has been
weighted with the beam particle fraction. In addition, we ignore the emission
from the jet cocoon and hot spot by limiting the calculations to values of the
beam particle fraction larger than 0.2 and to the inner $68 R_b$.

\subsection{Total and polarized emission as a function of the 
viewing angle}

  Because of the highly relativistic speeds in the jet, the emission is mainly
determined by the observing viewing angle, $\theta$, through the Doppler
factor and light aberration. Figures \ref{fig2} and \ref{fig3} show the
computed emission from the hydrodynamic model of A99 corresponding to a
viewing angle of 50$^{\circ}$ and 10$^{\circ}$, respectively. The emission is
computed for an optically thin observing frequency, and spectral index of the
electrons of 2.4.

  For relatively large viewing angles (Fig.\,\ref{fig2}) the jet emission is
limb brightened. This is in part due to the higher specific internal energy in
the shear layer, resulting in a larger synchrotron emission coefficient. On
the other hand, the Doppler factor can either enhance or cancel the limb
brightening depending on the value of the viewing angle $\theta$. Because of
the jet velocity stratification (see Fig. \ref{fig1}), for relatively large
viewing angles the fast jet spine suffers a larger amount of dimming than the
shear layer, enhancing the limb brightening. For our jet model, with a mean
$\Gamma\sim 7$ in the jet spine, this effect is maximized for $\theta\sim
50^{\circ}$, for which the shear layer emission is boosted while the jet spine
is dimmed (see the panel with the Doppler factor in Fig.\,\ref{fig2}). Cross
section profiles of the jet emission at different viewing angles are plotted
in Fig.\,\ref{fig4}, where the limb brightening effect can be observed more
easily. For small viewing angles, as corresponding to Fig.\,\ref{fig3}, the
jet spine emission is boosted, while the shear layer emission appears
dimmed. Details of the jet spine can be observed, as for instance two
recollimation shocks located at 26$R_b$ and 50$R_b$. The jet emission then
becomes spine brightened, instead of limb brightened, as observed in
Figs.\,\ref{fig3} and \ref{fig4}.

  The same arguments apply to the polarized flux. As a result of the helical
field in the shear layer, the apparent orientation of the magnetic field at
the jet edges is parallel to the jet axis. For the jet spine, the toroidal and
radial components of the magnetic field yield a net polarization perpendicular
to the jet axis. As shown in Figs.\,\ref{fig2} and \ref{fig4}, for relatively
large angles the aligned component of the helical magnetic field in the shear
layer projects into the jet spine partially canceling its field, yielding a
smaller net polarization, thereby stressing the limb brightening. Rails of low
polarization can be observed where the apparent magnetic field rotates between
being parallel (in the shear layer) to being perpendicular to the jet axis, as
observed in 3C~353 (Swain, Bridle \& Baum \cite{SB98}).

  Some of the kinematic and physical properties of the jet can be deduced by
analyzing the jet/counter jet emission ratio, plotted in Fig.\,\ref{fig5} for
the jet model of Fig.\,\ref{fig2}. The jet deceleration is apparent from a
progressively decrease in the total flux ratio along the jet axis. The
velocity stratification across the jet is also visible as a decrease of the
flux ratio close to the jet edges, that is, in the shear layer. This is
visible in the inner jet region, while further down the jet, when the jet
spine and shear layer velocities are more similar (due to the jet
deceleration), the jet/counter jet flux ratio is more uniformly distributed
across the jet. The slower velocity in the shear layer and its high emission
coefficient result in a smaller integrated flux ratio between the jet and
counter jet than for the case of a ``naked'' high velocity jet spine (see also
Komissarov \cite{K90}). This is because the shear layer emission is less
affected by the viewing angle through the Doppler factor.

\subsection{Jet cross section emission asymmetry}

  Because of the helical magnetic field structure in the shear layer, an
asymmetry in the emission appears across the jet. This asymmetry is more
pronounced in the polarized emission, and is a function of the viewing angle,
as shown in Fig.\,\ref{fig4}. In order to understand this effect we need to
study the variation across the jet of the angle between the magnetic field and
the line of sight {\it in the fluid frame}, $\vartheta$. The synchrotron
radiation coefficients are a function of the sine of this angle, and
asymmetries in the distribution of $\vartheta$ will be translated into the
emission maps. In order to compute $\vartheta$ we need to Lorentz transform
the line of sight from the observer's to the fluid's frame (see e.g., Rybicki
\& Lightman \cite{RL79})
\[
\sin\theta'=\frac{\sin\theta}{\Gamma(1-\beta\cos\theta)} \quad , \quad
\cos\theta'=\frac{\cos\theta-\beta}{(1-\beta\cos\theta)}
\]
where $\theta'$ is the viewing angle in the fluid frame. Consider a helical
magnetic field with a pitch angle $\phi$, measured with respect to the jet
axis. The angles $\vartheta^t$ and $\vartheta^b$ (where superscripts $t$ and
$b$ refer to the top and bottom of the jet, respectively) add $2 \phi$ (note
that $\vartheta^{t,b}$ is always defined as positive). Therefore, as long as
$\phi$ is different from zero or $\pi/2$, i.e. the field is neither purely
aligned nor toroidal, the factor $\sin\vartheta^{t,b}$ in the synchrotron
radiation coefficients will introduce an asymmetry in the jet emission. This
asymmetry will reach a maximum value for a helical magnetic field with
$\phi=\pi/4$, as the one considered here. However, independently of the helix
pitch angle, the predominance between $\sin\vartheta^t$ and $\sin\vartheta^b$
will reverse at $\theta'=\pi/2$, which corresponds to a viewing angle in the
observer's frame of $\cos\theta_r=\beta$. For a helical field oriented
clockwise as seen in the direction of flow motion (i.e., the aligned component
of the field is parallel to the jet flow), and for $\theta'<\pi/2$ the bottom
of the jet will show larger emission, while for $\theta'>\pi/2$ the top of the
jet will be brighter (the opposite is true for a helical field oriented
counter-clockwise, i.e. $\phi > \pi/2$). The maximum asymmetry will be
obtained for $\theta'=\phi$ and $\theta'=\pi-\phi$, and the fastest transition
(with changing $\theta'$) between top/bottom emission predominance will be
obtained for $\phi$ close to $\pi/2$, i.e. when little aligned field is
present.

  In the model we are considering, the shear layer has a mean $\Gamma\sim
1.7$, and therefore $\theta_r\sim 36^{\circ}$. Smaller angles will show bottom
jet dominance in emission, while for larger values the top of the jet will
appear brighter. This is more clearly visible in Fig.\,\ref{fig4}. Note also
that for the counter jet the helical field rotates opposite to the main jet,
and therefore the jet asymmetry emission reverses. This is particularly well
observed in the plot of the polarized emission ratio between the jet and
counter jet of Fig.\,\ref{fig5}.

\begin{figure*}
\plotone{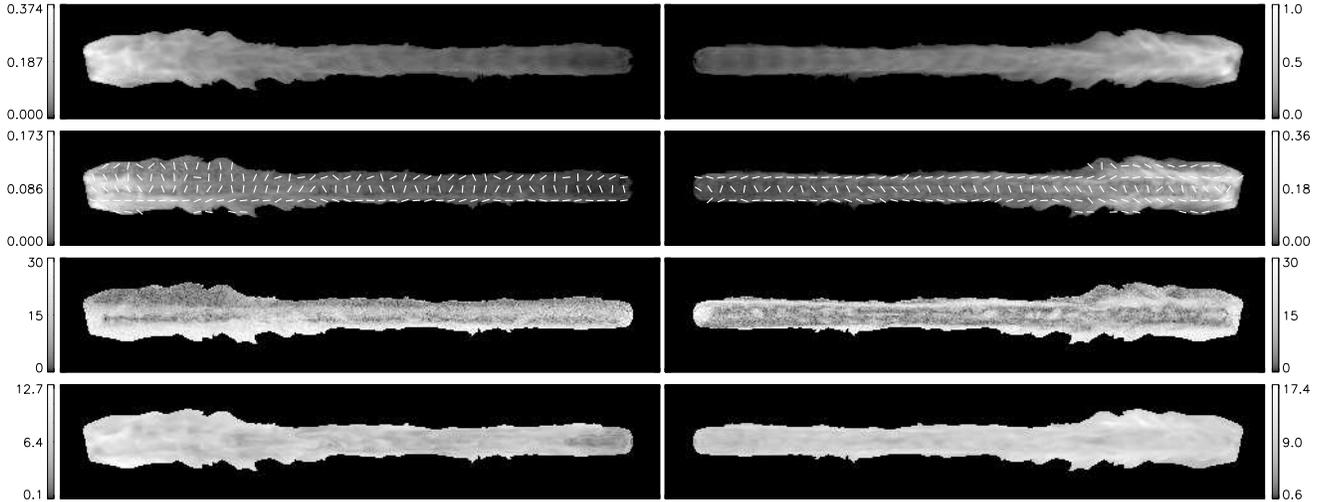}
%\vspace{7cm}
\caption{From top to bottom the panels show the total intensity, the polarized
intensity, the degree of polarization, and mean Doppler factor for a jet
viewed at an angle of 50$^{\circ}$ (right panels) and its counter jet (left
panels). Averages along the line of sight for each pixel, using the emission
coefficient as a weight, have been used to plot the Doppler factor. The total
and polarized intensities (in units normalized to the maximum of the main jet
total intensity) are plotted on a square root scale. The bars in the polarized
intensity panels show the direction of the magnetic field.}
\label{fig2}
\end{figure*}

\begin{figure*}[t]
\plotone{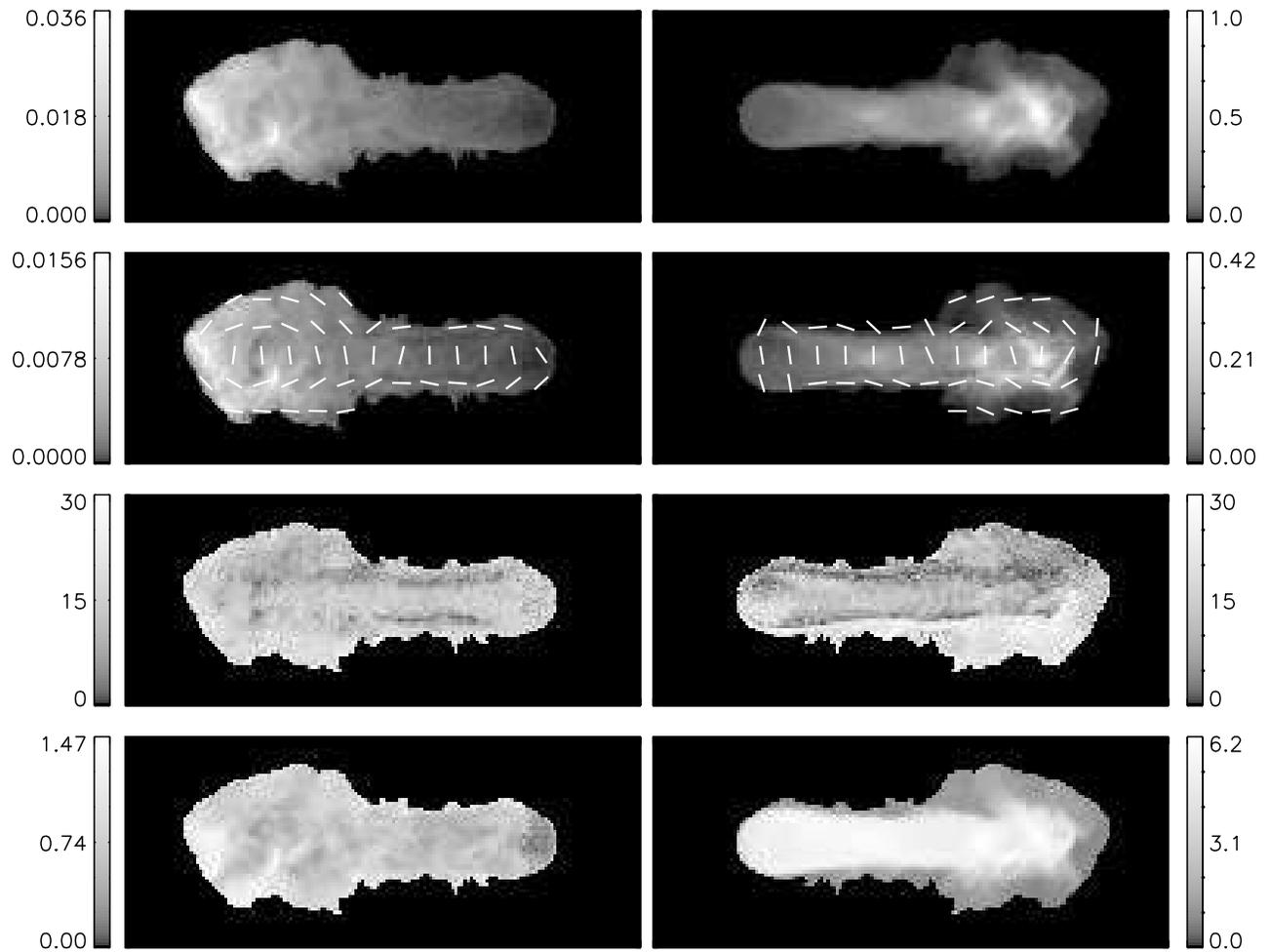}
%\vspace{3.5cm}
\caption{Same as Fig. \ref{fig2}, but for a viewing angle of 10$^{\circ}$.}
\label{fig3}
\end{figure*}

\begin{figure*}[t]
\plotone{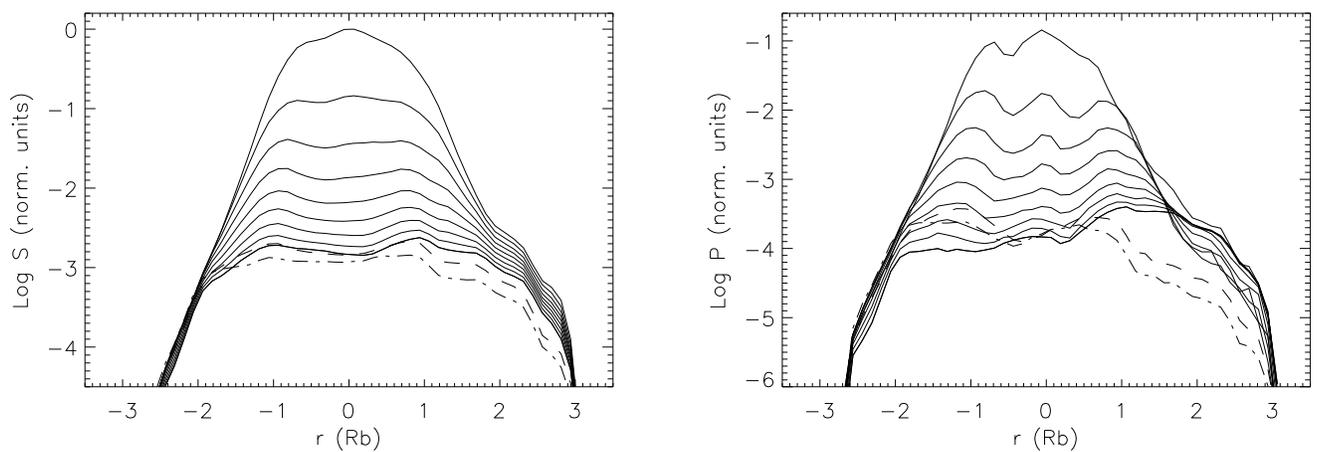}
%\vspace{6cm}
\caption{Logarithm of the integrated total ({\it left}) and polarized ({\it
right}) intensity across the jet for different viewing angles. Lines are
plotted in intervals of 10$^{\circ}$ between an angle of 10$^{\circ}$ (top
line in both plots) and 90$^{\circ}$ (showing a progressive decrease in
emission). Dashed (dot dashed) lines correspond to an observing angle of
-130$^{\circ}$ (-170$^{\circ}$). Positive beam radii correspond to the top in
the images of Figs.\,\ref{fig2} and \ref{fig3}. Units are normalized to the
maximum total intensity.}
\label{fig4}
\end{figure*}

\begin{figure*}[t]
\plotone{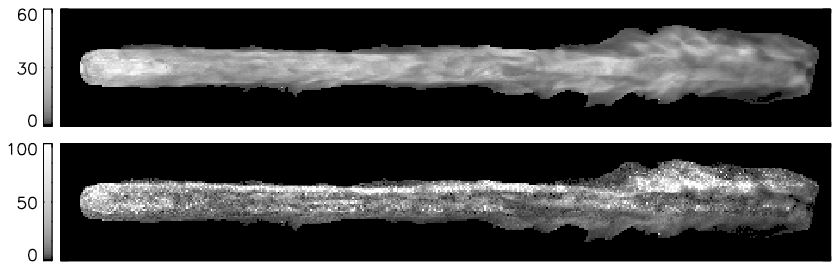}
%\vspace{1.5cm}
\caption{Total ({\it top}) and polarized ({\it bottom}) intensity jet/counter
jet ratios for the jet models of Fig.\,\ref{fig2}. The polarized ratio is
saturated at 100.}
\label{fig5}
\end{figure*}

  Although the $\sin\vartheta$ factor affects both the total and the polarized
emission, the asymmetry is more clearly present in the polarized flux (see
Figs.\,\ref{fig2}, \ref{fig4} and \ref{fig5}).  This is due to: i) The
presence of a randomly oriented magnetic field component, which renders the
magnetic field distribution more homogeneous in the jet and diminishes the
asymmetry. ii) Smaller values of $\vartheta$, independently whether present at
the top or the bottom of the jet, always represent a larger variation of the
magnetic field orientation along the line of sight. In practice this
represents a larger degree of randomness in the magnetic field along the
integration columns, decreasing the net polarization.

  It is interesting to note that for $\theta \sim \theta_r$, small changes in
the jet velocity or the viewing angle will produce a flip in the top/bottom
jet emission dominance. For fast jets, $\theta_r$ will be accordingly small,
and we will be biased towards observing jets with top emission predominance
(as long as the helical field rotates clockwise as seen in the direction of
flow motion).
  
  An interpretation of the polarization observations of the blazar 1055+018 by
Attridge, Roberts \& Wardle (\cite{ARW99}) can be obtained in terms of the
model presented here. For that, we need to assume that 1055+018 is oriented
close to $\theta_r$, and contains a shear layer with a helical field. If the
helical field is oriented clockwise, the polarized emission observed at the
top of the jet in inner regions would require that initially $\theta >
\theta_r$, or $\theta' > \pi/2$. To obtain the opposite situation further down
the jet, $\theta'$ has to become smaller than $\pi/2$, and for that either
$\theta$ decreases, or $\theta_r$ increases, which requires that $\beta$
decreases. A third less plausible possibility is that the helical field in the
shear layer changes orientation, i.e. the pitch angles becomes larger that
$\pi/2$. Therefore, we can successfully explain the flip in the top/bottom
orientation of the polarization asymmetry in 1055+018 if the jet bends towards
the observer, or if it decelerates. Attridge, Roberts \& Wardle (\cite{ARW99},
and references therein) report the existence of bends in the jet of
1055+018. This supports our hypothesis, but at the location of the flip in the
polarization emission asymmetry the jet spine emission decreases abruptly,
contrary to what would be expected in the case of a bend towards the observer
which should increase the jet spine emission by differential Doppler
boosting. Attridge, Roberts \& Wardle (\cite{ARW99}) obtained significantly
larger apparent velocities for components closer to the core suggesting a
deceleration along the jet. Therefore, this suggests our hypothesis of jet
deceleration as the most plausible for the sudden change in the
polarization predominance between the top and bottom of the jet in 1055+018,
since a jet deceleration will decrease the Doppler boosting, and hence the jet
spine emission as observed. A relatively small aligned field (helical pitch
angle close to $\pi/2$) will help to obtain such a fast flip in the
polarization asymmetry with a relatively small jet deceleration.

\begin{acknowledgements}
This research was supported by Spain's Direcci\'on General de Ense\~nanza
Superior (DGES) grants PB97-1164 and PB97-1432. MAA expresses his gratitude to
the Conselleria d'Educaci\'o i Ci\`encia de la Generalitat Valenciana for a
research fellowship. We thank A. Alberdi for comments that improved the
manuscript.
\end{acknowledgements}

\newpage

\begin{thebibliography}{1998}

\bibitem[1999a]{MA99a}Aloy, M. A., Ib\'a\~nez, J. M., Mart\'{\i}, J. M.,
G\'omez, J. L., \& M\"uller, E. 1999a, ApJ, 523, L125

\bibitem[1999b]{MA99b}Aloy, M. A., Ib\'a\~nez, J. M., Mart\'{\i}, J. M.,
\& M\"uller, E. 1999b, ApJS, 122, 151

\bibitem[1999]{ARW99}Attridge, J. M., Roberts, D. H., \& Wardle,
J. F. C. 1999, ApJ, 518, L87

\bibitem[1995]{JL95}G\'omez, J. L., Mart\'{\i}, J. M., Marscher, A. P.,
Ib\'a\~nez, J. M., \& Marcaide, J. M. 1995, ApJ, 449, L19

\bibitem[1997]{JL97}G\'omez, J. L., Mart\'{\i}, J. M., Marscher, A. P.,
Ib\'a\~nez, J. M., \& Alberdi, A. 1997, ApJ, 482, L33

\bibitem[1998a]{JL98a}G\'omez, J. L., Marscher, A. P., Alberdi, A.,
Mart\'{\i}, J. M., \& Ib\'a\~nez, J. M. 1998a, ApJ, 499, 221

\bibitem[1998b]{JL98b}G\'omez, J. L., Marscher, A. P., Alberdi, A.,
Mart\'{\i}, J. M., Ib\'a\~nez, J. M., \& Marchenko, S. G. 1998b, in ASP
   Conf. 159, BL Lac Phenomenon, ed. L. O. Takalo, \& A. Sillanp\"a\"a (San
Francisco: ASP), 435

\bibitem[1999]{JL99}G\'omez, J. L., Marscher, A. P., \& Alberdi, A. 1999, ApJ,
522, 74

\bibitem[1990]{K90}Komissarov, S. S. 1990. Sov. Astron. Lett. 16(4), 284

\bibitem[1996]{KF96}Komissarov, S. S., \& Falle, S. A. E. G. 1996, in ASP
Conf. 100, Energy Transport in Radio Galaxies and Quasars, ed. P. E. Hardee,
A. H. Bridle, \& J. A. Zensus (San Francisco: ASP), 165

\bibitem[1997]{KF97}Komissarov, S. S., \& Falle, S. A. E. G. 1997, MNRAS, 288,
833

\bibitem[1996]{L96}Laing, R. A. 1996, in ASP Conf. 100, Energy Transport in
Radio Galaxies and Quasars, ed. P. E. Hardee, A. H. Bridle, \& J. A. Zensus
(San Francisco: ASP), 241

\bibitem[1999]{L99}Laing, R. A., Parma, P., de Ruiter, H. R., \& Fanti,
R. 1999, MNRAS, 306, 513

\bibitem[1997]{MI97}Mioduszewski, A. J., Hughes, P. A., \& Duncan, G. C. 1997,
ApJ, 476, 649

\bibitem[1979]{RL79}Rybicki, G., Lightman, A. 1979. Radiative Processes in
Astrophysics. Wiley, New York, p. 110

\bibitem[1989]{SPA89}Sol, H., Pelletier, G., \& Asseo, E. 1989, MNRAS, 237,
411 

\bibitem[1998]{SB98}Swain, M. R., Bridle, A. H., \& Baum, S. A. 1998, ApJ,
507, L29

%\bibitem[1997]{T97}Zensus, J. A. 1997, Annu. Rev. Astron. Astrophys, 35, 607

\end{thebibliography}
\end{document}